\documentclass[a4paper,11pt]{article}
\usepackage{jinstpub} % for details on the use of the package, please see the JINST-author-manual
\usepackage{lineno}
\usepackage{subcaption}
\usepackage{tikz}
\title{\boldmath Application of exhaustive simulation flow for advanced performance prediction of monolithic active pixel sensors}

\author[h,1]{E. Sacchetti \note{Corresponding author.}}

\author[a]{M.~Babeluk}
\author[a]{T.~Bergauer}
\author[a]{M.~Friedl}
\author[a]{C.~Irmler}
\author[a]{B.~Pilsl}
\author[a]{R.~Russo}
\author[a]{C.~Schwanda}

\affiliation[a]{Marietta Blau Institute for Particle Physics, Austrian Academy of Sciences, Dominikanerbastei 16, 1010 Vienna, Austria}
\author[b]{L.~Gaioni}
\author[b]{V.~Re}
\author[b]{E.~Riceputi}
\author[b]{G.~Traversi}

\affiliation[b]{Department of Engineering and Applied Sciences, University of Bergamo, Viale Marconi 5, I-24044 Dalmine (BG), Italy}

\author[c]{S.~Giroletti}
\author[c]{L.~Ratti}

\affiliation[c]{Department of Electrical, Computer and Biomedical Engineering, University of Pavia, Via Ferrata 5, I-27100 Pavia, Italy}

\author[d,e]{G.~F.~Benfratello}
\author[d,e]{S.~Bettarini}
\author[e]{F.~Bosi}
\author[d,e]{G.~Casarosa}
\author[e]{L.~Corona}
\author[d,e]{F.~Forti}
\author[e]{A.~Gabrielli}
\author[e]{M.~Massa}
\author[d,e]{L.~Massaccesi}
\author[e]{M.~Minuti}
\author[e]{A.~Moggi}
\author[d,e]{S.~Mondal}
\author[d,e]{G.~Rizzo}
\author[e]{M.~Rovini}
\author[e]{A.~Taffara}

\affiliation[d]{Dipartimento di Fisica “E. Fermi”, Universit\`a di Pisa, L.go B. Pontecorvo 3, I-56127 Pisa, Italy}
\affiliation[e]{INFN Sezione di Pisa, L.go B. Pontecorvo 3, I-56127 Pisa, Italy}

\author[f]{M.~Barbero}
\author[f]{P.~Barrillon}
\author[f]{R.~Boudagga}
\author[f]{P.~Breugnon}
\author[f]{D.~Fougeron}
\author[f]{P.~Pangaud}
\author[f]{J.~Serrano}
\author[f]{V.~Vobbilisetti}
\author[f]{D.~Xu}

\affiliation[f]{Aix Marseille Univ, CNRS/IN2P3, CPPM, Marseille,  France}

\author[g]{D.~Auguste}
\author[g]{J.~Bonis}
\author[g]{Y.~Peinaud}
\author[g]{M.~Winter}

\affiliation[g]{Laboratoire de Physique des 2 infinis Ir\`ene Joliot-Curie – IJCLab, Université Paris-Saclay, CNRS/IN2P3, IJCLab, 91405 Orsay, France}

\author[h]{J.~Baudot}
\author[h]{G.~Bertolone}
\author[h]{A.~Dorokhov}
\author[h]{G.~Dujany}
\author[h]{L.~Federici}
\author[h]{C.~Finck}
\author[h]{A.~Himmi}
\author[h]{C.~Hu-Guo}
\author[h]{M. Kachel}
\author[h]{A.~Kumar}
\author[h]{M.~Maushart}
\author[h]{F.~Morel}
\author[h]{H.~Pham}
\author[h]{I.~Ripp-Baudot}
\author[h]{R.~Sefri}
\author[h]{P.~Stavroulakis}
\author[h]{I.~Valin}

\affiliation[h]{Universit\'e de Strasbourg, CNRS, IPHC UMR 7178, F-67000 Strasbourg, France}

\author[i]{F.~Bernlochner}
\author[i]{C.~Bespin}
\author[i]{J.~Dingfelder}
\author[i]{T.~Kishishita}
\author[i]{H.~Krüger}
\author[i]{L.~Schall}
\author[i]{M.~Vogt}

\affiliation[i]{Physikalisches Institut, Rheinische Friedrich-Wilhelms-Universität Universität Bonn, Nussallee 12, 53115 Bonn, Germany}

\author[j]{M.~Karagounis}

\affiliation[j]{University of Applied Sciences and Arts Dortmund, Sonnenstraße 96-100, 44139 Dortmund, Germany}

\author[k]{Y.~Buch}
\author[k]{A.~Frey}
\author[k]{B.~Schwenker}
\author[k]{M.~Schwickardi}

\affiliation[k]{II. Physikalisches Institut, Georg-August-Universität Göttingen, Friedrich-Hund-Platz 1, 37077 Göttingen, Germany}

\author[l,m]{K.~Hara}
\author[l,m]{D.~Jeans}
\author[l,m]{K.~R.~Nakamura}
\author[l,m]{Y.~Okazaki}

\affiliation[l]{High Energy Accelerator Research Organization (KEK), Tsukuba 305-0801, Japan}
\affiliation[m]{The Graduate University for Advanced Studies (SOKENDAI), Hayama 240-0193, Japan}

\author[n]{T.~Higuchi}

\affiliation[n]{Kavli Institute for the Physics and Mathematics of the Universe (WPI), University of Tokyo, Kashiwa-no-ha 5-1-5, Kashiwa 277-8583, Japan}

\author[o]{Y.~Onuki}
\author[o]{S.~Wang}

\affiliation[o]{Department of Physics, University of Tokyo, Hongo 7-3-1, Tokyo 113-0033, Japan}

\author[p]{C.~Lacasta}
\author[p]{C.~Marinas}
\author[p]{J.~Mazorra de Cos}
\author[p]{L.~Molina-Bueno}

\affiliation[p]{Instituto de Fisica Corpuscular (IFIC), CSIC-UV, Catedratico Jose Beltran, 2. E-46980 Paterna, Spain}

\author[q]{A.~Bevan}
\author[q]{M.~Bona}
\author[q]{D.~Howgill}

\affiliation[q]{School of Physical and Chemical Sciences, Department of Physics and Astronomy, Queen Mary University of London, 327 Mile End Road, London, E1 4NS, United Kingdom}

\author[r]{W.~ Song}
\author[r]{J.~Gong}
\author[r]{X.~Gao}

\affiliation[r]{College of Physics, Jilin University
, 2699 Qianjin Street, Changchun, Jilin, China}

\author[s]{A.~Fernandez~Prieto}
\author[s]{A.~Gallas~Torreira}

\affiliation[s]{Universidade de Santiago de Compostela, 2010 Instituto Galego de Física de Altas Enerxías (IGFAE), Colexio de San Xerome, PZ Obradoiro, S/N. E-15782 Santiago de Compostela, Spain }
\emailAdd{elio.sacchetti@iphc.cnrs.fr}

\abstract{
Monolithic active pixel sensor (MAPS) developments have pushed the detection performance in various directions, especially relative to timing where nanosecond-level precision is now considered.
This evolution calls for a simultaneous upgrade of the simulation tools.
We have developed a simulation flow that covers steps from the signal creation in the sensitive volume to the output of the pixel digital logic that performs the time-of-arrival and time-over-threshold (ToA/ToT) measurements.
This approach adds several new features to the traditional use the of the TCAD - Allpix Squared duo, among which : the integration of the pixel wells from the layout in order to precisely describe the pixel key characteristics such as leakage and punch-through currents and the coupling of Monte Carlo simulations (Allpix Squared) with high precision electrical simulations (SPICE). The first (Allpix Squared) for the precise description of the current induced at the collection electrode and the second (SPICE) to guarantee high precision simulation of the front-end electronics using realistic signal events. Irradiation is also modeled, both from the charge propagation side (charge trapping) and from the front-end response side (high input signal discharge).\\

\noindent We have applied this methodology to the MAPS developed in the context of the Belle II vertex detector upgrade. In this contribution, we detail the key features of the exhaustive simulation flow, present the outcome of the comparison with the TJ-Monopix2 measurements and discuss the interest of the methodology for the development of modern MAPS.

}


\begin{document}

\maketitle
\flushbottom

\section{Introduction}
\label{sec:intro}

Recent monolithic active pixel sensors (MAPS) developments~\cite{snoeys_monolithic_2023} have pushed the detection performance in various directions, especially relative to timing where nanosecond-level precision is now consider.
This evolution requires at the same time an upgrade of the simulation tools and methodologies used to develop these sensors, both on physics and integrated circuit design. In this contribution, we present and use the proposed simulation flow, from precise modeling of the 3D pixel model, irradiation modeling both from physics side and electrical side, up to reproducing accurate calibration measurements (before and after irradiation) using recently developed Monte~Carlo coupled simulation within the Allpix Squared framework, allowing to precisely describe all the steps of the signal processing. First of all, we describe the integration of the n and p-wells in the pixel 3D model, then we present the currents evolution with irradiation, by comparing the simulated results to measurements. Afterwards, the front-end response degradation with irradiation is studied, before reproducing in simulation the measurements performed with $\mathrm{^{55}Fe}$ radioactive source, by applying the proposed exhaustive simulation flow.\\
This work is presented in the context of the upgrade of the SuperKEK-B collider (Tsukuba, Japan), hosting the Belle II experiment for which a new fully pixelated vertex detector is being developed – the VTX. All the VTX detection layers will be equipped with the same MAPS: OBELIX (Optimized BELLE II pIXel sensor). Operation at room temperature after a NIEL fluence of 5~$\times$~$10^{14}$~1~MeV~$\mathrm{n_{eq}/cm^{2}}$ and providing time-stamping at 50~ns as baseline or 3~ns with increased power dissipation are key specifications of OBELIX. This chip is derived from the TJ-Monopix2 prototype ~\cite{rizzo_dmaps_2025,bespin_development_2022}. The TJ-Monopix2 prototype \cite{moustakas_thesis} was originally developed for ATLAS ITk, and is manufactured in the TowerJazz 180~nm CMOS modified process~\cite{Snoeys-2017}, with a low dose n-type implant intended to uniformize the depletion. The sensor is a 512x512 pixels matrix, with a 33.04~\textmu m pixel pitch. Several versions have been fabricated, and a single sensor also contains several pixel front-end flavors. The measurements and simulations presented in this work correspond to the version featuring these characteristics : 30~\textmu m epitaxial layer, "DC-cascode" analogue front-end. The pixel circuit features a time-of-arrival and time-over-threshold (ToA/ToT) logic. The ToT logic has 7 bits, with a 25~ns clock period. The TJ-Monopix2 prototype introduces an asymmetry in the deep p-well layer (shown in figure~\ref{fig:pixel_layout}), to enhance the lateral electric drift in the surrounding area. This feature is called RDPW (Reduced Deep P-Well), as opposed to the typical FDPW (Full Deep P-well). The presented simulations are compared with measurements performed on the three following proton-irradiated sensors: W8R4, W2R17 and W8R6, irradiated respectively at fluences up to: $\mathrm{\Phi}$~=~1, 2.5 and 5~$\times$~$10^{14}$~1~MeV~$\mathrm{n_{eq}/cm^{2}}$\cite{benfra_thesis}. The samples have undergone a long-term annealing of around one year between the irradiation and the measurements. There is an approximately 23~\% doping difference in the low dose n-type implant: W8 samples are "low" and W2 is "high".

\section{Layout integration in the pixel 3D model}
\label{sec:I_layout_integration}

\begin{figure}[ht]
    \centering
    \begin{subfigure}[b]{0.35\textwidth}
        \includegraphics[trim=0cm 0cm 0cm 0cm, clip, width=\textwidth]{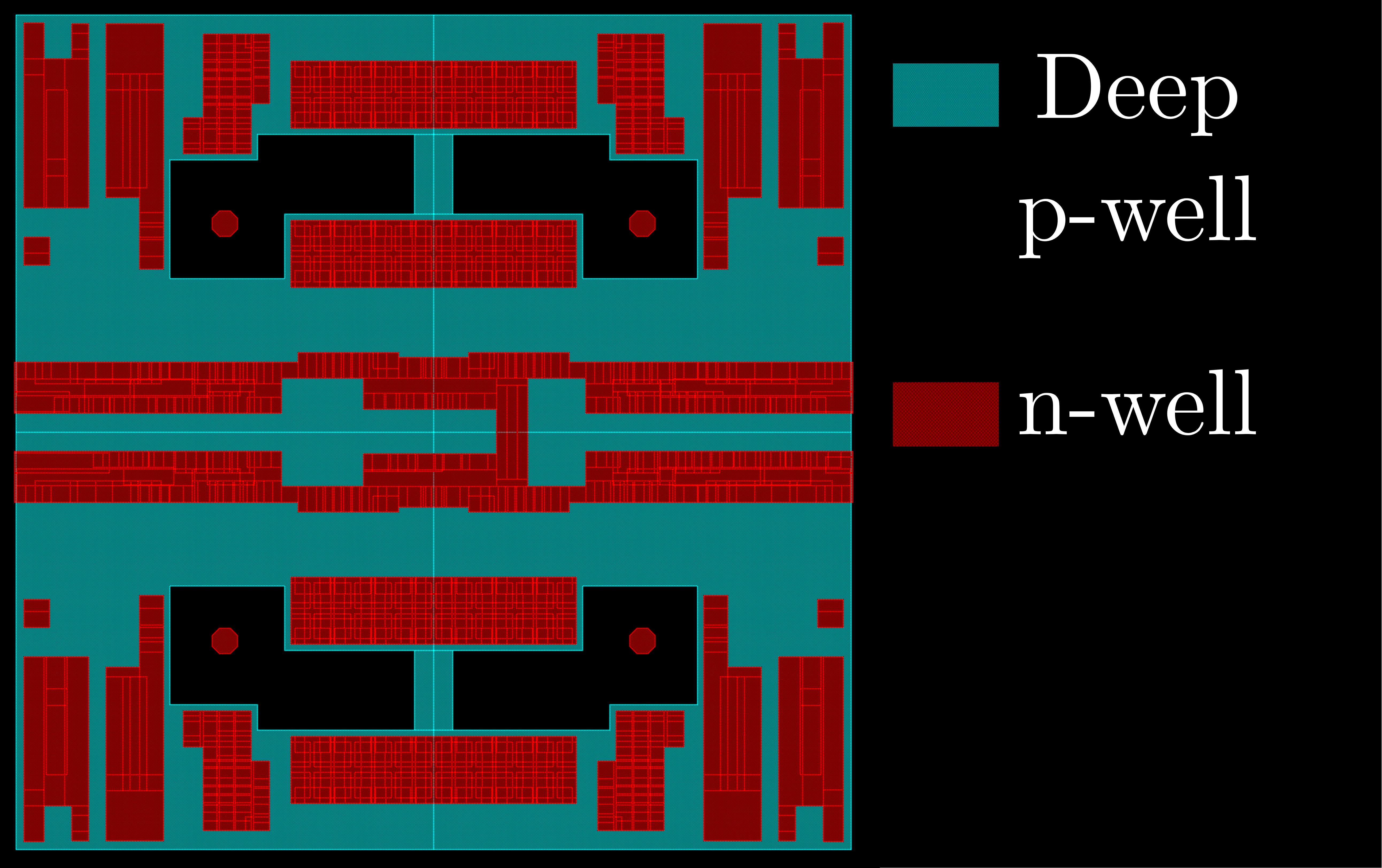}
        \caption{~}
        \label{fig:pixel_layout}
    \end{subfigure}
    % \hfill
    \begin{subfigure}[b]{0.3\textwidth}
        \includegraphics[trim=0cm 0cm 0cm 0cm, clip, width=\textwidth]{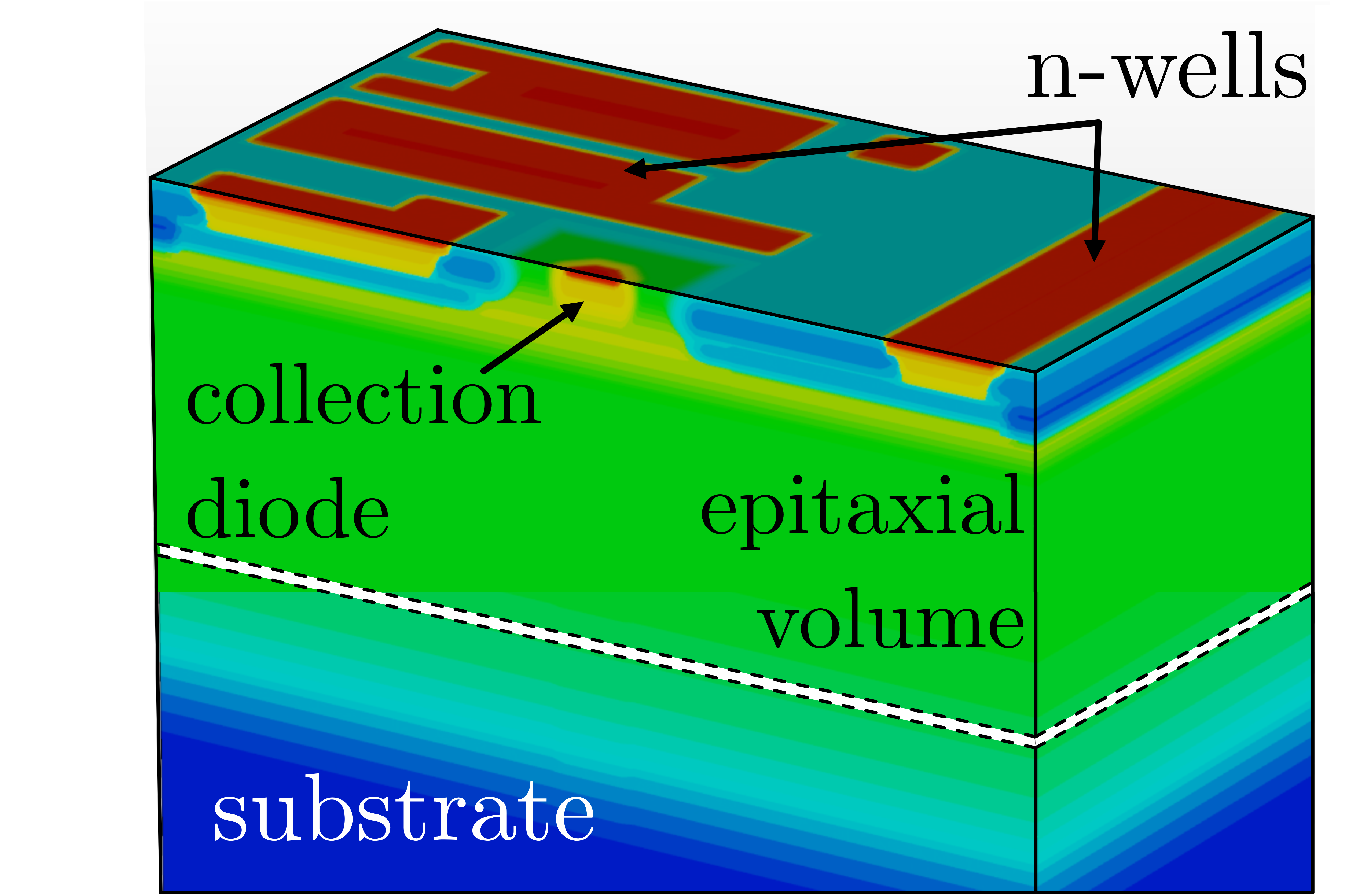}
        \caption{~}
        \label{fig:TCAD_pixel_3D}
    \end{subfigure}
    \vspace{-10pt}
    \caption{(a) 2x2 pixels layout,  (b) Cross-section model of a single pixel (not at scale).}
    \label{fig:2_pixel_layout}
\end{figure}

As introduced in section~\ref{sec:intro} and illustrated in figure~\ref{fig:pixel_layout}, the TJ-Monopix2 prototype features an asymmetric deep p-well layer in the pixel area, that was introduced to enhance the lateral electric field, and consequently has an effect on some pixel properties, mainly the charge collection efficiency.

\begin{figure}[ht]
    \centering
    \begin{subfigure}[c]{0.38\textwidth}
        \includegraphics[trim=0cm 0cm 0cm 0cm, clip, width=\textwidth]{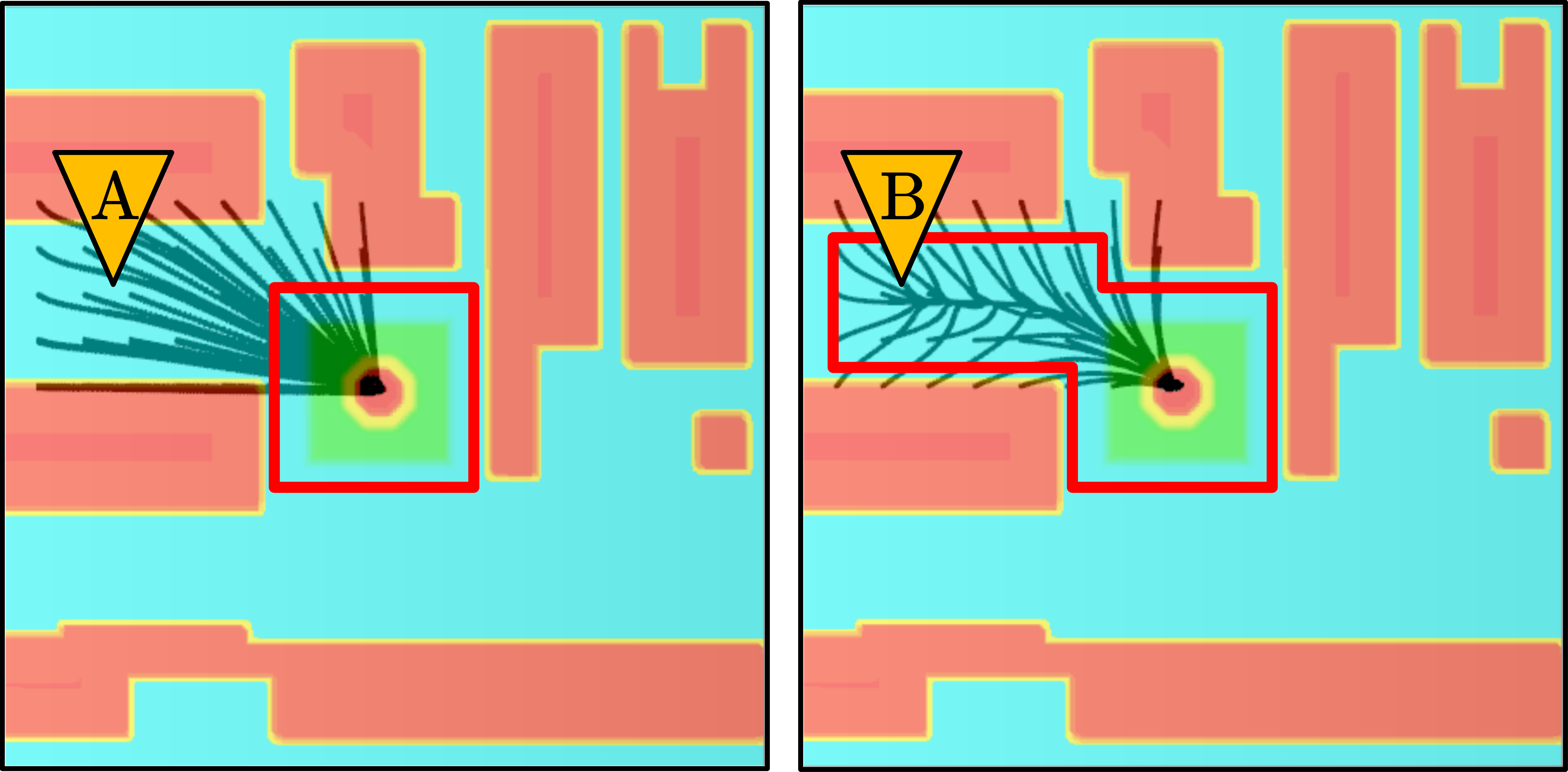}
        \caption{~}
        \label{fig:EF_streamlines}
    \end{subfigure}
    \hspace{20pt}
    \begin{subfigure}[c]{0.38\textwidth}
        \includegraphics[trim=0cm 0cm 0cm 0cm, clip, width=\textwidth]{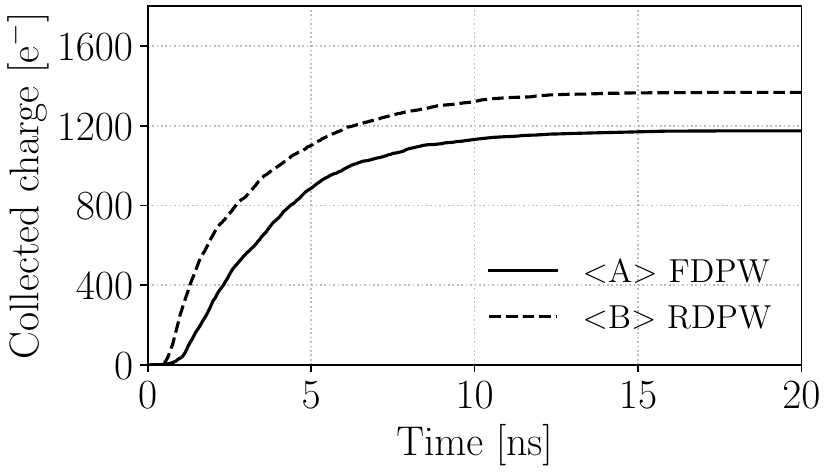}
        \caption{~}
        \label{fig:charge_collection_RDPW}
    \end{subfigure}
    \vspace{-10pt}
    \caption{(a) Top views of the pixel, FDPW (left) and RDPW (right) versions. The red polygons represent deep p-well's shape. The electric field streamlines are shown for the region of interest. A and B markers point the same position where the charge deposition is performed, for the two versions. (b) Charge collection profiles for the two versions, before irradiation.}
    \label{fig:2_charge_collection_RDPW}
\end{figure}

The regions where large area n-wells are placed have an impact on the in-pixel charge collection efficiency\cite{moustakas_thesis}.
An impinging charged particle crossing the sensing volume will deposit energy all along its path, generating electrons/holes pairs. Depending on its position, a portion of the electrons may be generated directly in the n-well, thus being collected by the n-well itself, without being propagated in the p-type epitaxial layer. Depending on the depth of the epitaxial layer, this loss of charge can correspond to a few percent of the total deposited charge. For example, considering a 30~\textmu m epitaxial depth, this loss can amount to approximately 4~\%. Taking into account these effects in simulation requires integration of the pixel layout in the 3D TCAD pixel model. To do so, the layers of interest are extracted from the GDS (Graphic Design System) files to the TCAD Sentaurus Structure Device Editor (SDE) environment. Preliminary operations are mandatory to correctly prepare the layers of interest and to be able to manipulate them independently when creating the 3D pixel model. The resulting 3D model contains the following layers, crucial for the charge collection: deep p-well, p-well, n-wells, low-dose n-type implant, extra p-stop implants.  The 3D pixel model created is presented in figure~\ref{fig:TCAD_pixel_3D}. To verify the effect of the deep P-Well on the efficiency of the charge collection, we compare the charge collection profile for the two FDPW and RDPW versions (figure~\ref{fig:2_charge_collection_RDPW}). In figure~\ref{fig:EF_streamlines}, the streamlines in black illustrate the difference in the electric field lines between the typical FDPW case (left) and the RDPW case (right). 

The collection diode voltage is set to +1.15~V, the p-well and the substrate voltages are set to -6~V. A charge deposition of 63~$\mathrm{e^-}$/\textmu m (assumed to correspond to the charge deposited by a Minimum Ionizing Particle (MIP) in a 30~\textmu m silicon thickness) is done through the sensor depth, for cases A and B, corresponding to FDPW and RDPW versions.  Their charge collection profiles (seed pixel), before irradiation, are presented in figure~\ref{fig:charge_collection_RDPW}. With the FDPW, the collected charge is 1175~$\mathrm{e^-}$ with a rise-time (defined as the duration between 10~\% and 90~\% of the total charge collection time) $\mathrm{t_{rise,FDPW}=6.14~ns}$, while the RDPW improves the collected charge up to 1365~$\mathrm{e^-}$ ($\approx$~16~\% better), with a rise-time $\mathrm{t_{rise,RDPW}=5.88~ns}$.\\
Including the discussed deep p-well asymmetry into the 3D TCAD model of the TJ-Monopix2 allows to increase the accuracy of the simulations, thanks to better modeling of the electric field in the pixel epitaxial layer. 
This 3D pixel model is used in the following sections to perform I-V simulations (section~\ref{sec:II_current_evolution}) and Monte Carlo simulations (section~\ref{sec:IV_coupled_MC_sims}).

\section{Currents evolution with irradiation}
\label{sec:II_current_evolution}

\begin{figure}[ht]
    \centering
    \begin{subfigure}[b]{0.4\textwidth}
        \includegraphics[width=\textwidth]{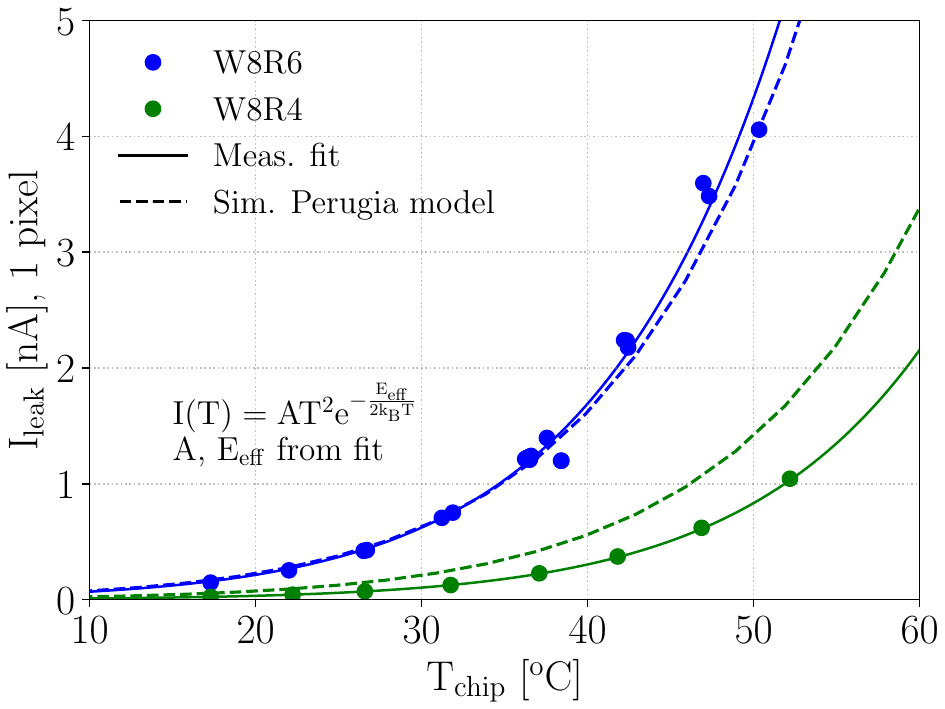}
        \caption{~}
        \label{fig:Ileak}
    \end{subfigure}
    \hspace{20pt}
    \begin{subfigure}[b]{0.4\textwidth}
        \includegraphics[width=\textwidth]{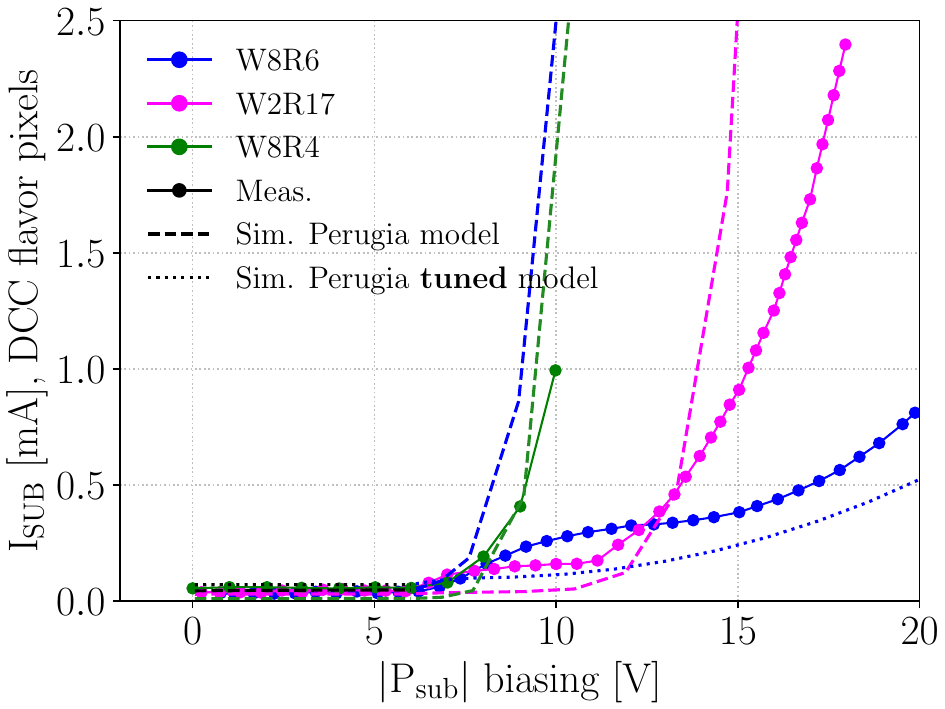}
        \caption{~}
        \label{fig:Ipt}
    \end{subfigure}
    \vspace{-10pt}
    \caption{(a) $\mathrm{I_{leak}}$, meas. and sim. (Perugia model), for different temperatures, PSUB set to -20~V, (b) $\mathrm{I_{pt}}$, meas. and sim., for different SUB bias voltages. PWELL set to -6~V in both cases.}
    \label{fig:side_by_side}
\end{figure}

The currents that impact performance of the pixel the most are the ones between the following electrodes: the substrate (SUB), the p-wells (PWELL) and the collection diode. The SUB and PWELL potentials are set at -6~V (before irradiation), while the CCE is biased by a reset diode at approximately $\mathrm{V_{CCE}}$~=~1.15~V. The current contributions to be studied are the leakage current and the punch-through current. The leakage current $\mathrm{I_{leak}}$, flowing towards the collection diode increases with irradiation and temperature. Characterization of $\mathrm{I_{leak}}$ is mandatory, as this requires modification of the DC-operating point of the front-end input node, as detailed in section~\ref{sec:III_FE_degradation}. The punch-through current, $\mathrm{I_{pt}}$ flows between SUB and PWELL when those electrodes are biased at different voltages. This current contribution decreases with irradiation (the resistance decreases). Before irradiation, measurements performed on the TJ-Monopix2 prototype have highlighted the impossibility to operate at a high (in absolute value) SUB voltage, while after NIEL irradiation it was possible to do so, due to effective build up of a positive charge at the edge of the PWELL, as in the double junction effect. TCAD simulations have been carried out to reproduce the I-V curves obtained in measurements with the samples W8R4, W2R17 and W8R6 (irradiation levels given in section \ref{sec:intro}). To include in the simulation the effects of irradiation on the evolution of currents, the University of Perugia model \cite{perugia_model} is used. We compare in figure~\ref{fig:side_by_side} the results of the measurements and simulations. In figure~\ref{fig:Ileak}, we give the evolution of $\mathrm{I_{leak}}$ in function of the temperature, given for both W8R4 and W8R6 samples.

We find that the simulations are in good agreement with the measurements. The leakage current behavior is well reproduced between 0.1~nA and 0.5~nA, especially for the most irradiated sample, W8R6. The radiation damage constant $\alpha$ is used as a figure of merit to compare the measurements with the simulations. It is defined by: $\alpha=I_{leak}/(\Phi_{eq}\times Vol)$ where we consider the leakage current $I_{leak}$ (calculated at $\mathrm{T=20^\circ C}$), at the given fluence ${\Phi_{eq}}$ in the volume of the device $Vol$ (we consider a full depletion of the epitaxial layer). In table~\ref{tab:I_leak_coeff} are presented the calculated $\alpha$ values, for both measurements and simulations, showing a reasonable agreement between measurement and simulation, especially for the W8R6 sample.
\begin{table}[ht]
    \centering
    \caption{Pixel $\mathrm{I_{leak}}$, and radiation damage constants, from the figure~\ref{fig:Ileak} fits}
    \smallskip
    \label{tab:I_leak_coeff}
    \begin{tabular}{c||c|c|c|c}
        \hline
        NIEL & A ($\mathrm{I_{leak}@20^\circ C}$) & $\mathrm{E_{eff}}$ & $\mathrm{\alpha}$ meas.&$\mathrm{\alpha}$ sim.\\ ($\mathrm{1~MeV~n_{eq}/cm^{2}}$) ~ & (nA) & (eV) & \multicolumn{2}{c}{$\mathrm{(10^{-17}~A\cdot cm^{-1})}$}\\
        \hline
        $\mathrm{1\times10^{14}}$ & 0.033 $\pm$ 0.001 & 1.65 $\pm$ 0.10 & 1.02 $\pm$ 0.04 & 2.04\\
        $\mathrm{5\times10^{14}}$ & 0.214 $\pm$ 0.008 & 1.53 $\pm$ 0.03 & 1.31 $\pm$ 0.05 & 1.07\\
        \hline
    \end{tabular}
\end{table}
In figure~\ref{fig:Ipt} we introduce the evolution of $\mathrm{I_{pt}}$ depending on the SUB voltage. The punch-through current starts flowing between SUB and PWELL when the two electrodes are biased at a different voltage. We can notice that higher irradiation levels reduce this current, making possible operation at SUB voltages down to ~-20~V (for the W8R6 sample). However, there is a discrepancy when comparing the measured $\mathrm{I_{pt}}$ with the simulated one, suggesting that this specific case is slightly outside the scope of the Perugia model. Trying to understand the modeling of $\mathrm{I_{pt}}$ have pushed the work in the direction of adding ad-hoc modifications to the Perugia model. Further optimization of the model, to better describe the effect of the positive charge build at the PWELL edge, is ongoing. Nevertheless, it should be noted that the 23~\% relative ratio of the low-dose n-type implant doping, between the W8 and W2 samples, when taken into account in the TCAD pixel cell, reproduces the same trend as in measurements. The AC-coupled simulation is also performed to extract the detector capacitance: $\mathrm{C_d=3.5~fF}$ (TCAD), matching the value obtained from layout parasitic extraction. The pixel TCAD simulation results are in good agreement with the measurements performed on the irradiated samples. For the purpose of modeling the effects of irradiation on the analogue in-pixel front-end, the simulated values of leakage current and detector capacitance are reused in the forthcoming steps of this exhaustive simulation flow presentation.

\section{Front-end response degradation with irradiation}
\label{sec:III_FE_degradation}

\begin{figure}[ht]
    \centering
    \includegraphics[width=.8\textwidth]{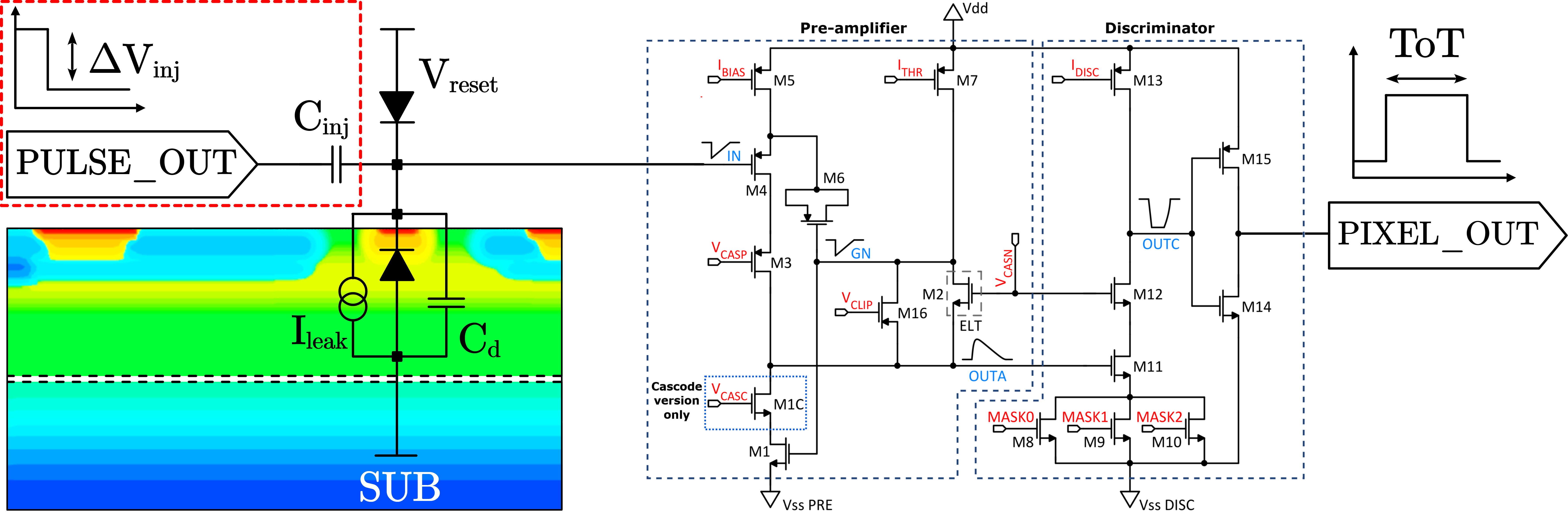}
    \caption{Pixel analogue circuit with the injection circuit (highlighted by the red dotted rectangle).}
    \label{fig:sch_front-end_model}
\end{figure}

The irradiation increases the leakage current flowing towards the front-end input, which causes a large sensor discharge. This effect significantly shorten the current pulse (due to a faster reset of the input) and in turn reduces the input signal amplitude.
We aim to reproduce the electrical behavior of the front-end in simulation, before and after irradiation. To do so, SPICE simulation of the front-end circuit is performed using Spectre (Cadence)~\cite{spectre}. We reproduce the same front-end bias conditions as in measurements. The effects of irradiation on the electrical behavior of the analogue pixel is modeled with a current source, representing the leakage current $\mathrm{I_{leak}}$, as presented in figure~\ref{fig:sch_front-end_model}.
Irradiation also causes surface damage (TID, Total Ionizing Dose) in the oxide layer ($\mathrm{SiO_2}$) by creating positive oxide charge (holes, being less mobile, can become trapped in the oxide). This can shift the threshold voltage of the transistors (negative shift for n-mos, positive for p-mos).

\begin{figure}[ht]
    \centering
    \begin{subfigure}[b]{0.4\textwidth}
        \includegraphics[trim=0cm 0cm 0cm 0cm, clip, width=1\textwidth]{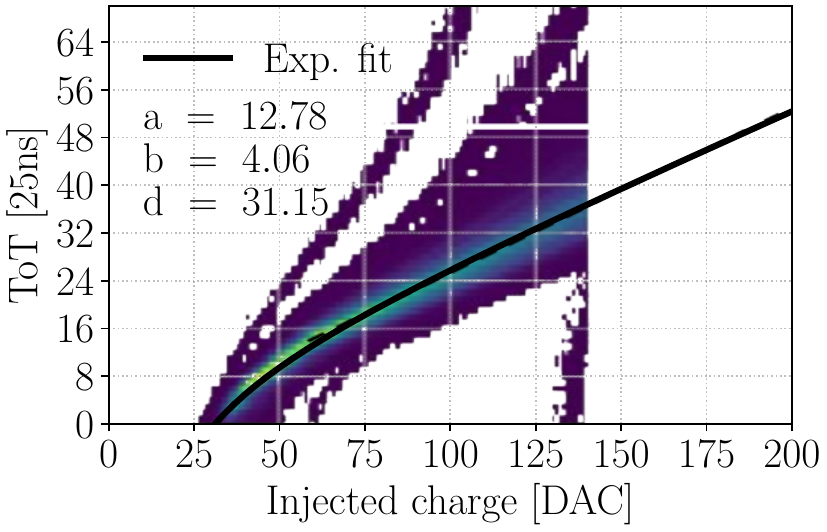}
        \caption{~}
        \label{fig:ToT_linearity_meas_before}
    \end{subfigure}
    \hspace{20pt}
    \begin{subfigure}[b]{0.4\textwidth}
        \includegraphics[trim=0cm 0cm 0cm 0cm, clip, width=1\textwidth]{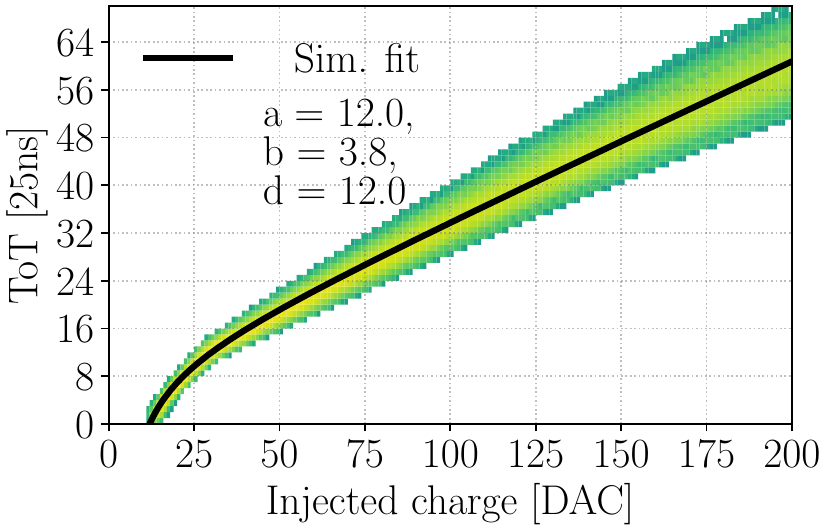}
        \caption{~}
        \label{fig:ToT_linearity_sim_before}
    \end{subfigure}

    \begin{subfigure}[b]{0.4\textwidth}
        \includegraphics[trim=0cm 0cm 0cm 0cm, clip, width=1\textwidth]{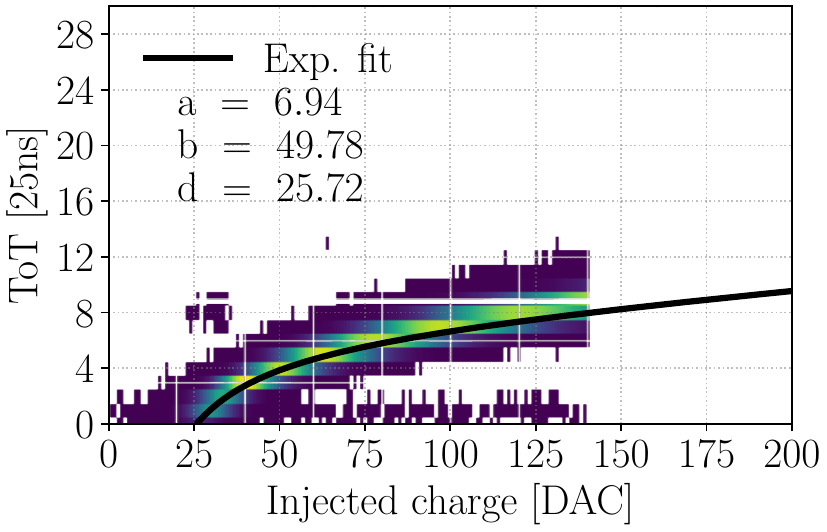}
        \caption{~}
        \label{fig:ToT_linearity_meas_after}
    \end{subfigure}
    \hspace{20pt}
    \begin{subfigure}[b]{0.4\textwidth}
        \includegraphics[trim=0cm 0cm 0cm 0cm, clip, width=1\textwidth]{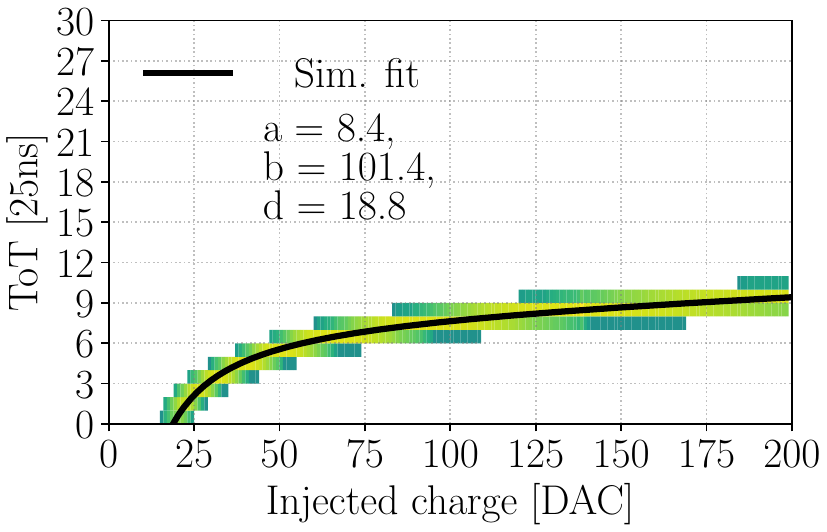}
        \caption{~}
        \label{fig:ToT_linearity_sim_after}
    \end{subfigure}
    \vspace{-10pt}
    \caption{ToT evolutions with injected charges, upper row: before irradiation, (a) measurement and (b) simulation, lower row: after irradiation, (c) measurement and (d) simulation}
    \label{fig:four_plots}
\end{figure}

We don't model the surface damage in our simulation, as the samples have been proton-irradiated (thus mainly subject to bulk damage), and due to a lack of foundry models describing surface damage effects on the transistor behavior. The in-pixel charge injection circuit implemented in TJ-Monopix2 is used for the internal ToT calibration (figure~\ref{fig:sch_front-end_model}), to emulate a signal corresponding to the one created by a particle hitting the pixel. The amount of injected charge can be controlled thanks to the on-chip 8-bit DAC, working in the range from 0 to 200 DAC units, with the smallest step of~7.03~mV. To reproduce accurately the pixel-to-pixel response dispersion, originating from transistor-to-transistor mismatch during the fabrication process, we use the Monte Carlo mismatch dispersion values provided by the foundry, available in the TJ-180 PDK (Process Design Kit). The simulation temperature is set according to the measured one, as is $\mathrm{I_{leak}}$: before irradiation, $\mathrm{T_{chip}=20~^\circ C}$, negligible leakage current ($\approx$ pA), and after irradiation, $\mathrm{T_{chip}=37~^\circ C}$, and $\mathrm{I_{leak}=1.2~nA}$. In figure~\ref{fig:four_plots} are presented the comparisons between the measured and simulated ToT curves, respectively before and after irradiation.
The following equation:~$\mathrm{ToT(Q_{inj})=(a/ Q_{inj}+b^{-1})(Q_{inj}-d)}$ is used to fit the ToT as function of the input charge. The fitted parameter values are given directly on the figure~\ref{fig:four_plots}. Simulations reproduce fairly the behavior observed in the measurements. Before irradiation, at a charge injected corresponding to 140~DAC (max. DAC value reachable in measurement), we obtain a ToT of 37 in measurements, and 45 in simulation. After irradiation, at 140~DAC, measurement gives a ToT of 8, and 10 in simulation. The quantitative agreement is within 20~\%, more importantly the effect of the high-sensor signal discharge is correctly modeled by integrating the leakage current into the electrical simulation after irradiation. It should be mentioned that the front-end DC-operating point is very sensitive to changes of the biasing currents, which makes it difficult to obtain the same exact result in simulation and in measurement.
This model of the front-end, including both the impact of the irradiation and the mismatch dispersion of its response will be now used to perform Monte Carlo coupled simulation.

\section{Coupling Monte Carlo physics and electrical simulations}
\label{sec:IV_coupled_MC_sims}

\begin{figure}[ht]
    \centering
    \begin{subfigure}[b]{0.32\textwidth}

        \begin{tikzpicture}
        \node[anchor=south west, inner sep=0] (image) at (0,0) {\includegraphics[width=\textwidth]{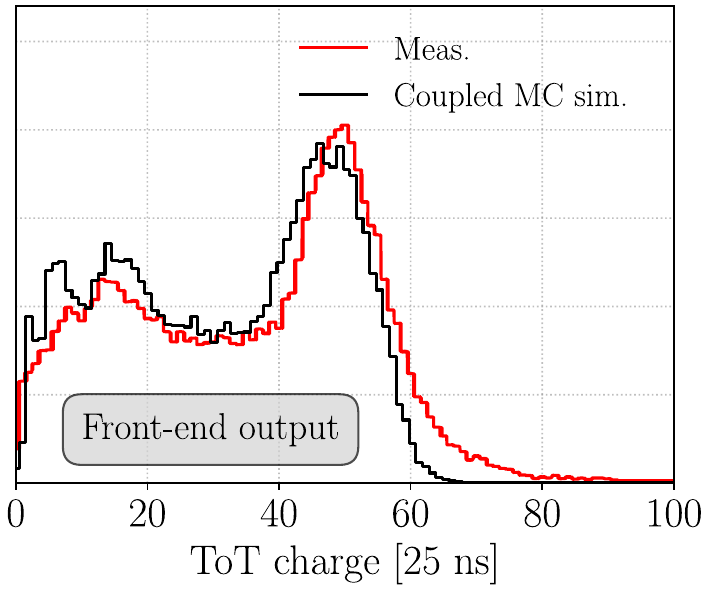}};
        \hspace{-5pt}
        \node[anchor=west, text width=0.3\textwidth, font=\small] at ([xshift=-3mm]image.west) {
            \rotatebox{90}{~~~~~~Before irradiation}
        };
        \end{tikzpicture}
        \caption{~}
        \label{fig:Fe55_before_PixelHit}
    \end{subfigure}
    \begin{subfigure}[b]{0.32\textwidth}
        \includegraphics[width=\textwidth]{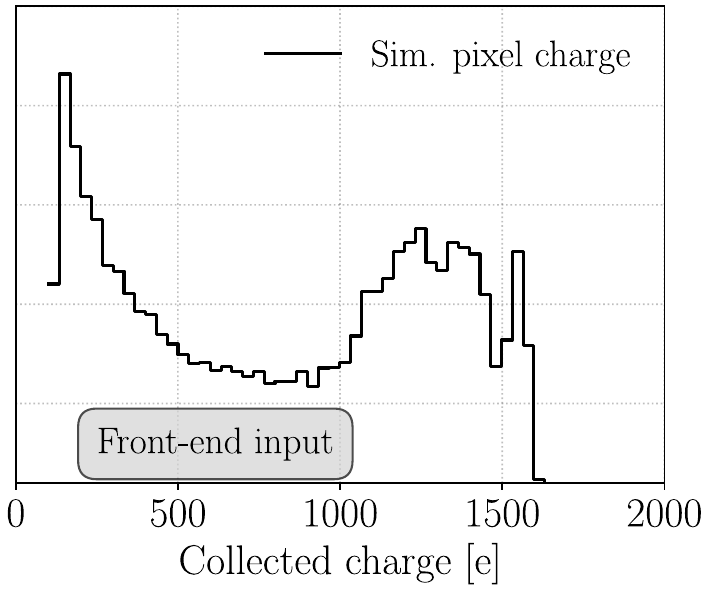}
        \caption{~}
        \label{fig:Fe55_before_PixelCharge}
    \end{subfigure}
    \hfill
    \begin{subfigure}[b]{0.32\textwidth}
        \includegraphics[width=\textwidth]{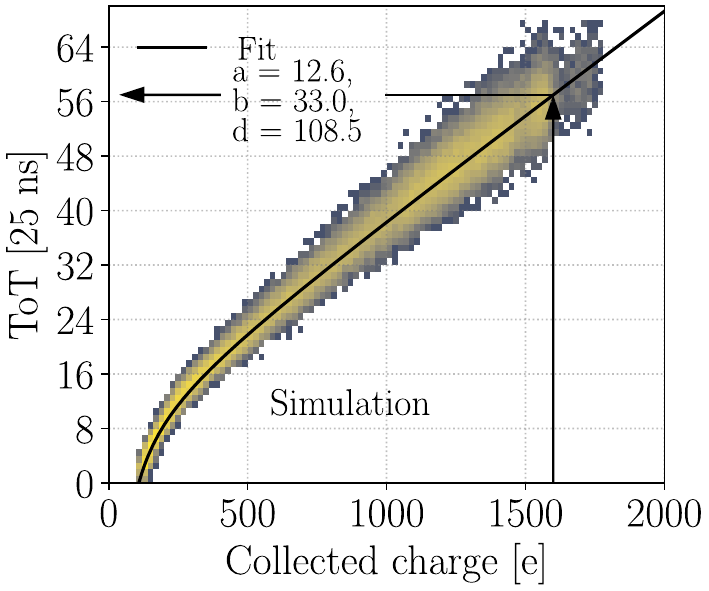}
        \caption{~}
        \label{fig:Fe55_before_2D_hist}
    \end{subfigure}
    \begin{subfigure}[b]{0.32\textwidth}

        \begin{tikzpicture}
        \node[anchor=south west, inner sep=0] (image) at (0,0) {\includegraphics[width=\textwidth]
        {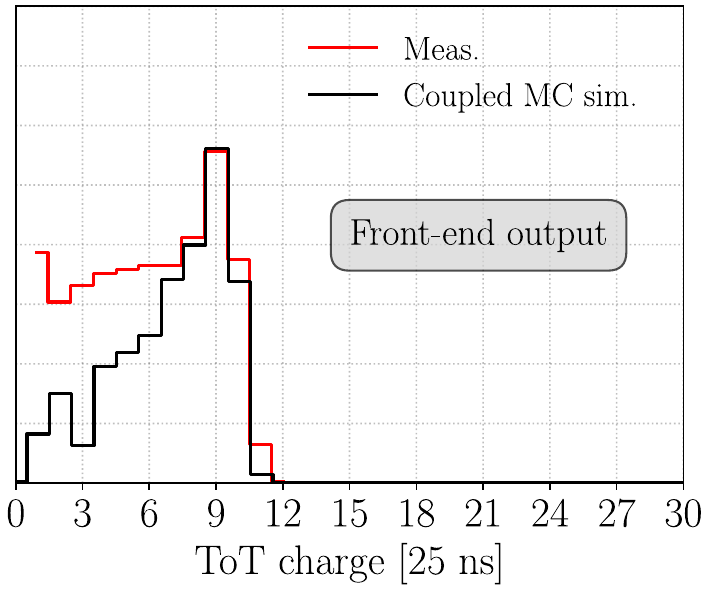}};
        \hspace{-5pt}
        \node[anchor=west, text width=0.3\textwidth, font=\small] at ([xshift=-3mm]image.west) {
            \rotatebox{90}{~~~~~~After irradiation}
        };
        \end{tikzpicture}
    
        \caption{~}
        \label{fig:Fe55_after_PixelHit}
    \end{subfigure}
    \hfill
    \begin{subfigure}[b]{0.32\textwidth}
        \includegraphics[width=\textwidth]{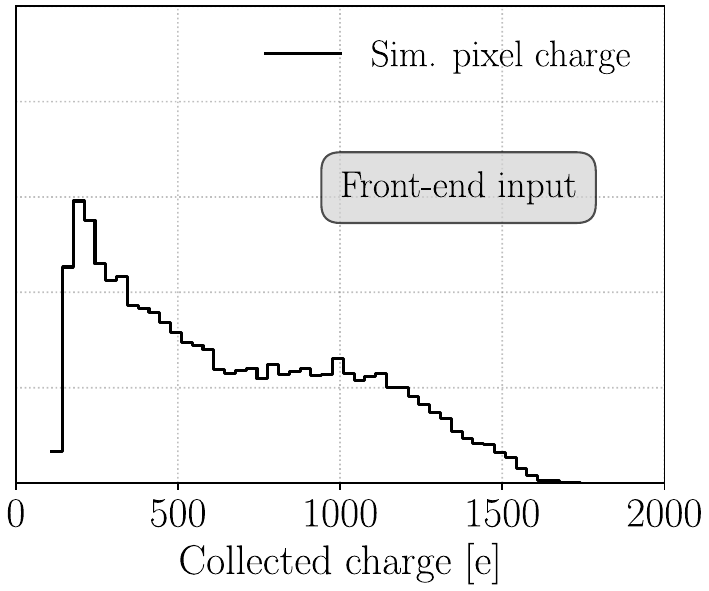}
        \caption{~}
        \label{fig:Fe55_after_PixelCharge}
    \end{subfigure}
    \hfill
    \begin{subfigure}[b]{0.32\textwidth}
        \includegraphics[width=\textwidth]{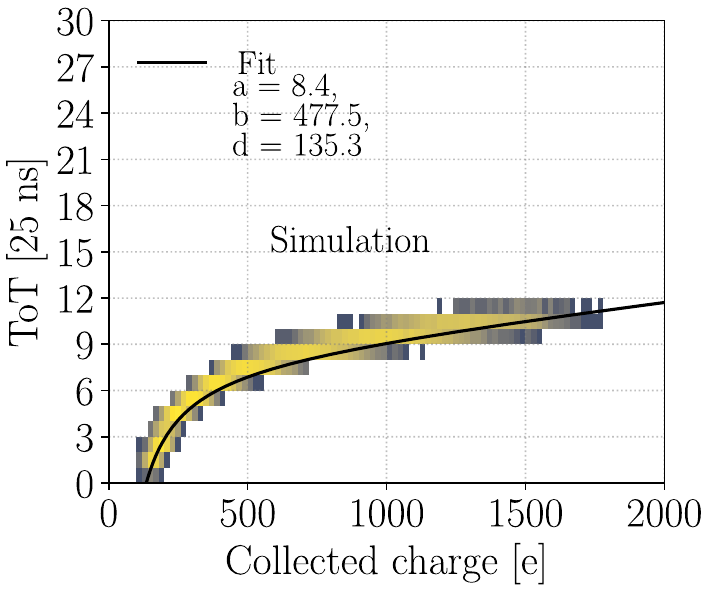}
        \caption{~}
        \label{fig:Fe55_after_2D_hist}
    \end{subfigure}
    \vspace{-10pt}
    \caption{(a), (b) and (c), before irradiation, respectively : the simulated and measured ToT histograms (front-end output), the simulated collected charge histogram (front-end input), and the 2D histogram of the simulated ToT depending on the simulated charge. (d), (e) and (f), after irradiation, same histograms}
    
    \label{fig:Fe55}
\end{figure}

Once the the effects of the irradiation have been characterized and modeled in TCAD and SPICE simulations, we have reproduced $\mathrm{^{55}Fe}$ measured spectra with TJ-Monopix2 prototype in simulation. The methodology consists of using the latest Allpix Squared~\cite{allpix} developments - the integration of SPICE simulation within the Allpix Squared Monte Carlo environment. The simulation of the deposition, the propagation and the collection of the charge is performed by Allpix Squared, whereas the electrical simulation of the front-end is done by SPICE, within a single event. The SPICE simulation is fully integrated within Allpix Squared via the \texttt{[NetlistWriter]} module, to benefit from the accuracy of both simulators with a user-friendly approach. It should be also mentioned, that each pixel is assigned a mismatch ID value (corresponds to the seed of the Monte Carlo simulation) in order to guarantee that the same pixel will have the same mismatch value from one event to another while pixel-to-pixel dispersion is still considered (introduced in section~\ref{sec:III_FE_degradation}). We compare the ToT histograms (front-end output), obtained in measurement and simulation, of a $\mathrm{^{55}Fe}$ calibration before and after irradiation in figures~\ref{fig:Fe55_before_PixelHit} and~\ref{fig:Fe55_after_PixelHit}.  As illustrated in these two figures, measurements results can be well reproduced in simulation.

With the simulation, the collected charge (raw front-end input, without considering the front-end discharge effect) can be accessed. We present these histograms, before and after irradiation, respectively in figures~\ref{fig:Fe55_before_PixelCharge} and~\ref{fig:Fe55_after_PixelCharge}. Before irradiation, one can distinguish the $\mathrm{^{55}Fe~K_\alpha}$ peak at $\approx$ 1600~e$^-$, contrary to after irradiation, due to the trapping of charges. 
The charge collected on the diode, and the resulting front-end analogue output (expressed in ToT with a 25~ns clock), both obtained in simulation, can be represented in a 2D histogram (figures~\ref{fig:Fe55_before_2D_hist} and~\ref{fig:Fe55_after_2D_hist}).
This permits to determine to which ToT value corresponds the $\mathrm{^{55}Fe~K_\alpha}$ peak ($\mathrm{ToT=57}$ before irradiation), and thanks to the ToT linearity curve presented in figure~\ref{fig:ToT_linearity_sim_before}, to calculate the injection capacitance value $\mathrm{C_{inj, sim}}$ following the same procedure as the one performed with the measurements.
This ToT value corresponds to $\mathrm{Inj_{^{55}Fe}=180~DAC}$, giving an injection capacitance of $\mathrm{C_{inj,sim}=1600/180=8.9~e^-/DAC}$, matching the one obtained with the measurements: $\mathrm{C_{inj, meas}=~9~e^-/DAC}$. The front-end loss of signal is a combination of trapping and sensor discharge caused by the high leakage current. By reaching a good matching between measurements and simulations of $\mathrm{^{55}Fe}$ source spectra, we confirm that the effects of the irradiation on the front-end behavior are correctly modeled in our coupled Monte Carlo simulation, emphasizing the interest of the presented exhaustive simulation flow.

\section{Conclusion}

We presented the main features of the proposed exhaustive simulation flow. The integration of the n/p-wells into the 3D TCAD pixel model helps to refine the electric field conditions of the sensing volume directly impacting the charge collection properties. The I-V simulations were performed to reproduce leakage and punch-through current measurements on irradiated sensors. Thanks to the Perugia model, we obtained a correct description of the evolution of the leakage current as a function of the temperature and irradiation levels. We modeled the effect of the irradiation on the front-end electrical behavior with a current source, reusing the simulated I-V results, which allowed to reproduce the sensor discharge similar to the measurements. By embedding SPICE simulation into Allpix Squared, we extended the framework to electrical simulations and successfully reproduced $^{55}$Fe calibrations both pre- and post-irradiation. By comparing measurements with simulations we validated the methodology and demonstrated the interest of the proposed simulation flow. This methodology aims to be applied to the ongoing sensor R\&D programs, and the future generations of MAPS.

\vspace{-3.4pt}
\acknowledgments

This work has received the support from the European Union’s Horizon 2020 Research and Innovation programme under Grant Agreements no 101004761 (AIDAinnova), Multilateral Scientific and Technological Cooperation in the Danube Region (MULT 03/23), Horizon 2020 ERC-Consolidator Grant No. 819127, TY-FJPPN (Toshiko Yuasa-France Japan Particle Physics Network), the MCIU with funding from the European Union NextGenerationEU (PRTR-C17.I01) and Generalitat Valenciana (GVANEXT), Project ASFAE/2022/016.

\vspace{-3.4pt}
\bibliographystyle{JHEP}
\bibliography{biblio.bib}

@article{snoeys_monolithic_2023,
    author = {W. Snoeys},
    title = "{Monolithic CMOS Sensors for high energy physics — Challenges and perspectives}",
    doi = "https://doi.org/10.1016/j.nima.2023.168678",
    journal = "Nucl. Instrum. Meth. A",
    volume = "1056",
    pages = "168678",
    year = "2023"
}

@article{rizzo_dmaps_2025,
	author = "Rizzo, G. and others",
	title = "{The DMAPS upgrade of the Belle II Vertex Detector}",
	doi = "10.1016/j.nima.2024.170164",
    journal = "Nucl. Instrum. Meth. A",
	volume = "1072",
	pages = "170164",
	year = "2025",
}

@article{bespin_development_2022,
	author = "Bespin, Christian and others",
	title = "{Development and characterization of a DMAPS chip in TowerJazz 180 nm technology for high radiation environments}",
	doi = "10.1016/j.nima.2022.167189",
    journal = "Nucl. Instrum. Meth. A",
    volume = "1040",
	pages = "167189",
	year = "2022",
	}

@phdthesis{moustakas_thesis,
      author = "Moustakas, Konstantinos",
      title = "{Design and Development of Depleted Monolithic Active Pixel Sensors with Small Collection Electrode for High Radiation Applications}",
      school = "Bonn U.",
      year = "2021",
      url = "https://cds.cern.ch/record/2782279",
      note = "Presented 09 Sep 2021",
}

@phdthesis{benfra_thesis,
    author = "Benfratello, Guglielmo F.",
    title = "{Characterization of irradiated monolithic CMOS active pixel sensors for the upgrade of the Belle II Vertex Detector}",
    school = "Pisa U.",
    year = "2025",
    url = "https://docs.belle2.org/pub_data/documents/4828/"
}

@article{perugia_model,
  author= "Moscatelli, F. and others",
  title="{Combined Bulk and Surface Radiation Damage Effects at Very High Fluences in Silicon Detectors: Measurements and TCAD Simulations}", 
  journal="IEEE TNS",
  doi="10.1109/TNS.2016.2599560",
  volume="63",
  pages="2716",
  year="2016"

}

@manual{spectre,
  title        = {{Spectre Circuit Simulator}},
  organization = {Cadence Design Systems, Inc.},
  edition      = {Version 23.1},
  year         = {2024},
  url          = {https://www.cadence.com}
}

@article{allpix,
    author = "Spannagel, S. and others",
    title = "{Allpix2: A modular simulation framework for silicon detectors}",
    journal = "Nucl. Instrum. Meth. A",
    doi="10.1016/j.nima.2018.06.020",
    year = "2018",
    volume="901",
    pages="164-172",
}

@article{Snoeys-2017,
    author = "Snoeys, W. and others",
    title = "{A process modification for CMOS monolithic active pixel sensors for enhanced depletion, timing performance and radiation tolerance}",
    doi = "10.1016/j.nima.2017.07.046",
    journal = "Nucl. Instrum. Meth. A",
    volume = "871",
    pages = "90--96",
    year = "2017"
}

\end{document}